\documentclass[11pt]{article}
\usepackage{amssymb}

\begin{document}

\title{ Higgs-Boson Production Associated with a Single Bottom Quark in Supersymmetric QCD  } \vspace{3mm}

\author{{ Hou Hong-Sheng$^{2}$, Ma Wen-Gan$^{1,2}$, Wu Peng$^{2}$, Wang Lei$^{2}$ and Zhang Ren-You$^{2}$}\\
{\small $^{1}$CAST (World Laboratory), P.O.Box 8730, Beijing, 100080, People's Republic of China} \\
{\small $^{2}$Department of Modern Physics, University of Science and Technology of China (USTC),}\\
{\small       Hefei, Anhui 230027, People's Republic of China} }

\date{}
\maketitle
\vskip 12mm

\begin{abstract}
Due to the enhancement of the couplings between Higgs boson and
bottom quarks in the minimal sypersymmetric standard model (MSSM),
the cross section of the process $pp(p\bar{p}) \to
h^0b(h^0\bar{b})+X$ at hadron colliders can be considerably
enhanced. We investigated the production of Higgs boson associated
with a single high-$p_T$ bottom quark via subprocess $bg(\bar{b}g)
\to h^0b(h^0\bar{b})$ at hadron colliders including the
next-to-leading order (NLO) QCD corrections in MSSM. We find that
the NLO QCD correction in the MSSM reaches $50\% \sim 70\%$ at the
LHC and $60\% \sim 85\%$ at the Fermilab Tevatron in our chosen
parameter space.
\par
\end{abstract}

\vskip 5cm


\vfill \eject

\baselineskip=0.32in

\renewcommand{\theequation}{\arabic{section}.\arabic{equation}}
\renewcommand{\thesection}{\Roman{section}.}
\newcommand{\nb}{\nonumber}

\makeatletter      
\@addtoreset{equation}{section}
\makeatother       

\par
\section{INTRODUCTION}
\par
One of the most important tasks of future high-energy experiments
is to search for scalar Higgs particles and investigate the
symmetry breaking mechanism of the electroweak $SU(2) \times
U(1)$. In the standard model (SM) \cite{sm}, one doublet of
complex scalar fields is needed to spontaneously break the
symmetry, leading to a single neutral Higgs boson $h^0$. But there
exists the problem of the quadratically divergent contributions to
the corrections to the Higgs boson mass . This is the so-called
naturalness problem of the SM. One of the good methods to solve
this problem is to make supersymmetric (SUSY) extensions to the
SM. Then the quadratic divergences of the Higgs mass can be
cancelled by loop diagrams involving the supersymmetric partners
of the SM particles exactly. The most attractive supersymmetric
extension of the SM is the minimal supersymmetric standard model
(MSSM) \cite{mssm-1,mssm-2}. In the MSSM, there are two Higgs
doublets $H_1$ and $H_2$ to give masses to up- and down-type
fermions. The Higgs sector consists of three neutral Higgs bosons,
one $CP$-odd particle ($A^0$), two $CP$-even particles ($h^0$ and
$H^0$), and a pair of charged Higgs bosons ($H^{\pm}$).
\par
In the SM, the so-called Yukawa couplings describe the
interactions between Higgs and fermions. Their coupling strengths
are proportional to $m_f/v$, where $v$ is the vacuum expectation
value of the Higgs field. Since the bottom quark mass is
approximately $5$ GeV, the Yukawa coupling to bottom quarks is
relatively weak. This leads to a small cross section for the
associated production of the Higgs boson ($h^0$) and bottom quarks
at hadron colliders \cite{willen1,willen2}. However, in the MSSM,
the Yukawa coupling between the Higgs boson ($h^0$) and bottom
quarks can be considerably enhanced for large values of
$\tan{\beta}=v_2/v_1$, where $v_1$ and $v_2$ are the vacuum
expectation values of the two Higgs boson fields $H_1$ and $H_2$,
respectively.
\par
Because the high-$p_T$ bottom quark can be tagged with reasonably
high efficiency, the observation of a bottom quark with high $p_T$
can reduce the backgrounds of the Higgs boson production. The
leading-order subprocess for production of the Higgs boson
associated with bottom quark is $bg \to bh^0$ \cite{zhu}.
Recently, the investigations of the process $pp(p\bar{p}) \to
bg(\bar{b}g) \to h^0b(h^0\bar{b})+X$ in the SM including the NLO
QCD corrections is presented in Ref. \cite{campbell1}. Their
calculation shows the cross section of the subprocess of $bg \to
h^0b$ is an order of magnitude larger than those of $gg(q \bar{q})
\to b \bar{b}h^0$ at the Fermilab Tevatron and the CERN LHC. They
find that the NLO QCD correction ranges from 50-60$\%$ at the
Tevatron for $m_h=100-200$ GeV, and at the LHC the correction with
$m_h=120-500~GeV$ ranges from $20-40 \%$ for $p_T>15~GeV$ and
$25-45 \%$ for $p_T>30~GeV$. They conclude that this production
mechanism improves the prospects for the discovery of a Higgs
boson with enhanced coupling to the bottom quarks.
\par
In this paper, we calculated the cross section for the associated
production of the Higgs boson and a single high-$p_T$ bottom quark
via $bg \to h^0b(\bar{b}g \to h^0\bar{b})$ in the MSSM at the
Tevatron and the LHC including the NLO QCD corrections. The
structure of this paper is as follow: In Sec. II, we discuss the
LO results of the subprocess $bg \to h^0b$. In Sec. III, we
present the calculations of the NLO QCD corrections. In Sec. IV,
the numerical results and conclusions are presented. Some lengthy
analytical expressions are listed in Appendices.

\section{THE LEADING ORDER CROSS SECTION}
Since the cross sections for the subprocess $bg \to h^0b$ and its
charge-conjugate subprocess $\bar{b}g \to h^0\bar{b}$ in the
CP-conserved MSSM are same, we present only the calculation of the
subprocess $b(p_1)g(p_2) \to h^0(k_3)b(k_4)$ here (where $p_{1,2}$
and $k_{3,4}$ represent the four-momentum of the incoming partons
and the outgoing particles, respectively.). The subprocess $bg \to
h^0b$ can occur through both s-channel and t-channel as shown in
Fig.1(A-B). So we divide the tree-level amplitude into two parts
and denote it as
\begin{eqnarray}
M^0=M_0^{(s)}+M_0^{(t)},
\end{eqnarray}
where $M_0^{(s)}$ and $M_0^{(t)}$ represent the amplitudes arising
from the s-channel diagram shown in Fig.1(A) and the t-channel
diagram shown in Fig.1(B) at the tree-level, respectively. The
explicit expressions for the amplitudes $M_0^{(s)}$ and
$M_0^{(t)}$ can be written as
\begin{eqnarray}
M_0^{(s)}&=& \frac{g_s(\mu_r) Y_b(\mu_r)}{\hat{s}}
   \bar{u}_i(k_4)( \rlap/p_1 + \rlap/p_2) \gamma_{\nu} u_j(p_1)\epsilon_{\nu}(p_2)T^a_{ij},  \nb \\
M_0^{(t)}&=& \frac{g_s(\mu_r) Y_b(\mu_r)}{\hat{t}}
   \bar{u}_i(k_4) \gamma_{\nu} (\rlap/p_1 - \rlap/k_3)
   u_j(p_1)\epsilon_{\nu}(p_2)T^a_{ij},
\end{eqnarray}
where $\hat{s}=(p_1+p_2)^2$, $\hat{t}=(p_1-k_3)^2$ and
$\hat{u}=(p_1-k_4)^2$ are the usual Mandelstam variables.
$g_s(\mu_r)$ is the running strong coupling strength and $T^a$ is
the $SU(3)$ color matrix. $Y_b(\mu_r)$ is the Yukawa coupling
between Higgs boson and bottom quarks. In MSSM, $Y_b(\mu_r)$ is
given as
\begin{eqnarray}
Y_b(\mu_r)=i \frac{g_w \overline{m}_b(\mu_r)}{2 m_W} \frac{{\rm
sin} \alpha}{{\rm cos} \beta},
\end{eqnarray}
where $\alpha$ is the mixing angle which leads to the physical
Higgs eigenstates $h^0$ and $H^0$. $\overline{m}_b(\mu_r)$ is the
$\overline{\rm MS}$ mass of the bottom quark. Throughout our
evaluation we adopt the simplified
Aivazis-Collins-Olness-Tung(ACOT) scheme\cite{aivazis}, which will
not make loss of accuracy in our calculation of subprocess $bg \to
h^0b$. That means the bottom quark mass is neglected except in the
Yukawa couplings during our calculation.
\par
Then the lowest order cross section for the subprocess $bg \to
h^0b$ in the MSSM is obtained by using the following formula:
\begin{eqnarray}
\label{folding} \hat{\sigma}^0(\hat{s}, bg \to h^0b) = \frac{1}{16
\pi \hat{s}^2} \int_{\hat{t}_{min}}^{\hat{t}_{max}} d\hat{t}~
\overline{\sum} |M^0|^2,
\end{eqnarray}
where $\hat{t}_{max}=0$ and $\hat{t}_{min}=m_h^2-\hat{s}$. The
summation is taken over the spins and colors of initial and final
states, and the bar over the summation recalls averaging over the
spins and colors of initial partons.

\section{NLO QCD CORRECTIONS}
\par
The NLO QCD contributions to the associated production of the
Higgs boson and a single bottom quark can be separated into two
parts: the virtual corrections arising from loop diagrams and the
real gluon emission corrections.
\par
\begin{flushleft} {\bf 1. Virtual Corrections } \end{flushleft}
\par
The virtual corrections in the MSSM to $bg \to h^0b$ consist of
self-energy, vertex and box diagrams which are shown in Figs.2-3.
Fig.2 shows the one-loop diagrams of the SM-like QCD corrections
from quarks and gluons, and Fig.3 presents the one-loop diagrams
of the SUSY QCD corrections from squarks and gluinos. There exist
both ultraviolet(UV) and soft/collinear infrared(IR) singularities
in the amplitude from the SM-like diagrams in Fig.2, and the
amplitude part from SUSY QCD diagrams(Fig.3) only contains UV
singularities. In our calculation, we adopt the 't Hooft-Feynman
gauge and all the divergences are regularized by using dimensional
regularization method in $d=4-2 \epsilon$ dimensions.
\par
In order to remove the UV divergences, we need to renormalize the
wave functions of the external fields, the strong coupling and the
$h^0-b-\bar{b}$ Yukawa coupling. For the renormalization of the
strong and Yukawa couplings, we employ the modified Minimal
Subtraction ($\overline{\rm MS}$) scheme. The relevant
renormalization constants in this work are defined as
\begin{eqnarray}
\label{defination of renormalization constants}
m_b & \to & m_b+\delta m_b,~~~~g_s \to (1+\delta g_s)g_s    \nb \\
Y_b & \to &Y_b+\delta Y_b,~~~~\delta Y_b=Y_b \frac{\delta
m_b}{m_b} \nb \\
b & \to & \left( 1+\frac{1}{2}\delta Z_{b}\right)b  \nb \\
g_{\mu} & \to & (1+ \frac{1}{2}\delta Z_g)g_{\mu},
\end{eqnarray}
where $g_s$ and $Y_b$ denote the strong coupling and the
$h-b-\bar{b}$ Yukawa coupling. $b$ and $g_{\mu}$ denote the fields
of bottom quark and gluon. The explicit expressions of these
renormalization constants are presented in Appendix A.
\par
Then the renormalized amplitude for virtual corrections, $M^{V}$,
can be divided into two parts and expressed as
\begin{eqnarray}
M^{V}=M^{loop}+M^{CT},
\end{eqnarray}
where $M^{loop}$ is the amplitude from one loop-diagrams shown in
Figs.2-3 and $M^{CT}$ is the amplitude from the diagrams which
contain counter-terms shown in Fig.4. The expression of the
$M^{loop}$ can be written in the form as
\begin{eqnarray}
\label{virtual diagrams amplitude}
 M^{loop} &=& \bar{u}_i(k_4) [ f_1
\gamma_{\mu} P_L+  f_2 \gamma_{\mu} P_R +
 f_3 p_{1\mu} P_L + f_4 p_{1\mu} P_R +
f_5 k_{4 \mu} P_L + f_6 k_{4 \mu} P_R  +
f_7 p_{1 \mu} \rlap/p_2 P_L    \nb \\
&+& f_8 p_{1 \mu} \rlap/p_2 P_R + f_9 k_{4 \mu} \rlap/p_2 P_L +
f_{10} k_{4 \mu} \rlap/p_2 P_R  + f_{11} \gamma_{\mu} \rlap/p_2
P_L + f_{12} \gamma_{\mu} \rlap/p_2 P_R
]u_j(p_1)\epsilon_{\mu}(p_2)T^a_{ij},
\end{eqnarray}
where $P_{L,R}=\frac{1 \mp \gamma_5}{2}$. The explicit expressions
of form factors $f_i~(i =1 \thicksim 12)$ are presented in
Appendix B, and $M^{CT}$ is expressed in Appendix A.
\par
The virtual corrections to the cross section can be written as
\begin{eqnarray}
\hat{\sigma}^{V}(\hat{s}, bg \to h^0b) = \frac{1}{16 \pi
\hat{s}^2} \int_{\hat{t}_{min}}^{\hat{t}_{max}} d\hat{t}~ 2 Re
\overline{\sum} [(M^{V})^{\dagger} M^0],
\end{eqnarray}
with $\hat{t}_{max}=0$ and $\hat{t}_{min}=m_h^2-\hat{s}$ and again
the summation with bar means the same operations as appeared in
Eq.(\ref{folding}).
\par
After the renormalization procedure, $\hat{\sigma}^{V}$ is
UV-finite. Nevertheless, it still contains the soft/collinear IR
singularities
\begin{eqnarray}
\label{virtual cross section}
d\hat{\sigma}^V|_{IR}=\left[\frac{\alpha_s}{2 \pi}
\frac{\Gamma(1-\epsilon)}{\Gamma(1-2 \epsilon)}\left(\frac{4 \pi
\mu_r^2}{\hat{s}}\right)^{\epsilon}\right]d\hat{\sigma}^0
\left(\frac{A^V_2}{\epsilon^2}+\frac{A^V_1}{\epsilon} \right),
\end{eqnarray}
where
\begin{eqnarray}
A^V_2&=&-\frac{17}{3}, \nb \\
A^V_1&=&-\frac{47}{6}+3 \ln \frac{-\hat{t}}{\hat{s}-m_{h}^2}
-\frac{1}{3} \ln \frac{-\hat{u}}{\hat{s}-m_{h}^2}.
\end{eqnarray}
The soft divergences can be cancelled by adding with the soft real
gluon emission corrections, and the remaining collinear
divergences are absorbed into the parton distribution functions,
which will be discussed in the following subsections.

\begin{flushleft} {\bf 2.  Real gluon emission Corrections } \end{flushleft}

The $O(\alpha_s)$ corrections to $bg \to h^0b$ due to real gluon
emission (shown in Fig.5) give the origin of IR singularities
which cancel exactly the analogous singularities present in the
$O(\alpha_s)$ virtual corrections mentioned in above subsection.
These singularities can be either of soft or collinear nature and
can be conveniently isolated by slicing the $bg \to h^0b+g$ phase
space into different regions defined by suitable cutoffs, a method
which goes under the general name of Phase Space Slicing(PPS). In
this paper, we have calculated the cross section for the $2 \to 3$
process
\begin{eqnarray}
b(p_1)+g(p_2) \to h^0(k_3)+b(k_4)+g(k_5),
\end{eqnarray}
using the method named two cutoff phase space slicing
method\cite{Harris}. We define the invariants
\begin{eqnarray}
\hat{s}&=&(p_1+p_2)^2,~~\hat{t}=(p_1-k_3)^2,~~\hat{u}=(p_1-k_4)^2,
\nb \\
\hat{s}_{45}&=&(k_4+k_5)^2,~~\hat{t}_{15}=(p_1-k_5)^2,~~
\hat{t}_{25}=(p_2-k_5)^2,~~\hat{t}_{45}=(k_4-k_5)^2,
\end{eqnarray}
and describe this method briefly as follows. Firstly, by
introducing an arbitrary small soft cutoff $\delta_s$ we separate
the $2 \to 3$ phase space into two regions, according to whether
the energy of the emitted gluon is soft, i.e. $E_5 \leq
\delta_s\sqrt{\hat{s}}/2$, or hard, i.e. $E_5 >
\delta_s\sqrt{\hat{s}}/2$. The partonic real cross section can be
written as
\begin{eqnarray}
\hat{\sigma}^R(bg \to h^0bg)=\hat{\sigma}^S(bg \to
h^0bg)+\hat{\sigma}^H(bg \to h^0bg),
\end{eqnarray}
where $\hat{\sigma}^S$ is obtained by integrating over the soft
region of the emitted gluon phase space, contains all the soft IR
singularities. Secondly, to isolate the remaining collinear
singularities from $\hat{\sigma}^H$, we further decompose
$\hat{\sigma}^H$ into a sum of hard-collinear (HC) and
hard-non-collinear ($\overline{\rm HC}$) terms by introducing
another cutoff $\delta_c$ named collinear cutoff
\begin{eqnarray}
\hat{\sigma}^H(bg \to h^0bg)=\hat{\sigma}^{\rm HC}(bg \to
h^0bg)+\hat{\sigma}^{\overline{\rm HC}}(bg \to h^0bg).
\end{eqnarray}
The HC regions of the phase space are those where the invariants
$t_{15},t_{25},t_{45}$ become smaller in magnitude than $\delta_c
\hat{s} $, in collinear condition, while at the same time the
emitted gluon remains hard. $\hat{\sigma}^{\rm HC}$ contains the
collinear divergences. In the soft and HC region, $\hat{\sigma}^S$
and $\hat{\sigma}^{\rm HC}$ can be obtained by performing the
phase space integration in $d$-dimension analytically. In the
$\overline{\rm HC}$ region, $\hat{\sigma}^{\overline{\rm HC}}$ is
finite and may be evaluated in four dimensions using standard
Monte Carlo techniques\cite{Lepage}. The cross sections,
$\hat{\sigma}^S$, $\hat{\sigma}^{\rm HC}$ and
$\hat{\sigma}^{\overline{\rm HC}}$, depend on the two arbitrary
parameters, $\delta_s$ and $\delta_c$. However, in the total real
gluon emission hadronic cross section $\hat{\sigma}^R$, after mass
factorization, the dependence on these arbitrary cutoffs cancels,
as will be explicitly shown in Sec. IV. This constitutes an
important check of our calculation. In the next two subsections,
we will discuss in detail the soft, hard-collinear gluon emission.

\begin{flushleft} {\bf 2.1  Soft gluon emission } \end{flushleft}

In the $p_1+p_2$ rest frame, the emitted gluon's $d$-momentum
($d=4-2 \epsilon$) can be parameterized as
\begin{eqnarray}
k_5=E_5(1,...,\sin \theta_1 \sin \theta_2,\sin \theta _1 \cos
\theta_2,\cos \theta_1)
\end{eqnarray}
The soft region of the $bg \to h^0b+g$ phase space is defined by
\begin{eqnarray}
0<E_5 \leq \delta_s\sqrt{\hat{s}}/2
\end{eqnarray}
In the soft region, the three body phase space can be factorized
as\cite{Harris}
\begin{eqnarray}
d\Gamma_3|_{soft}=d\Gamma_2 \left[\left(\frac{4
\pi}{\hat{s}}\right)^{\epsilon}
\frac{\Gamma(1-\epsilon)}{\Gamma(1-2 \epsilon)}\frac{1}{2(2
\pi)^2}\right] dS
\end{eqnarray}
with
\begin{eqnarray}
dS=\frac{1}{\pi}
\left(\frac{4}{\hat{s}}\right)^{-\epsilon}\int_0^{\delta_s
\sqrt{\hat{s}}/2 } dE_5 E_5^{1-2 \epsilon} \sin^{1-2 \epsilon}
\theta_1 d \theta_1 \sin^{-2 \epsilon}\theta_2 d\theta_2.
\end{eqnarray}
In the soft limit, the matrix element squared for the real gluon
emission, $\overline{\sum}|M^R|^2$, can be factorized into the
Born matrix element squared times an eikonal factor $\Phi_{eik}$
\begin{eqnarray}
\overline{\sum}|M^R(bg \to h^0b+g)|^2=(4 \pi \alpha_s \mu_r^{2
\epsilon}) \overline{\sum}|M^0(bg \to h^0b)|^2 \Phi_{eik},
\end{eqnarray}
where
\begin{eqnarray}
\Phi_{eik}= N \frac{p_1 \cdot p_2}{(p_1 \cdot k_5)(p_2 \cdot
k_5)}-\frac{1}{N}  \frac{p_1 \cdot k_4}{(p_1 \cdot k_5)(k_4 \cdot
k_5)}+N \frac{p_2 \cdot k_4}{(p_2 \cdot k_5)(k_4 \cdot k_5)},
\end{eqnarray}
with $N=3$. The partonic differential cross section in the soft
region can be written as
\begin{eqnarray}
d\hat{\sigma}^S=(4 \pi \alpha_s \mu_r^{2
\epsilon})\left[\left(\frac{4 \pi}{\hat{s}}\right)^{\epsilon}
\frac{\Gamma(1-\epsilon)}{\Gamma(1-2 \epsilon)}\frac{1}{2(2
\pi)^2}\right] d\hat{\sigma}^0 \int dS \Phi_{eik}.
\end{eqnarray}
After integration over the eikonal factors, the differential cross
section is
\begin{eqnarray}
\label{soft cross section} d\hat{\sigma}^S=d\hat{\sigma}^0
\left[\frac{\alpha_s}{2 \pi} \frac{\Gamma(1-\epsilon)}{\Gamma(1-2
\epsilon)}\left(\frac{4 \pi
\mu_r^2}{\hat{s}}\right)^{\epsilon}\right]
\left(\frac{A^S_2}{\epsilon^2}+\frac{A^S_1}{\epsilon}+A^S_0
\right),
\end{eqnarray}
with
\begin{eqnarray}
A^S_2&=&\frac{17}{3}, \nb \\
A^S_1&=&-\frac{34}{3} \ln \delta_s - 3 \ln
\frac{-\hat{t}}{\hat{s}-m_{h}^2}
+\frac{1}{3} \ln \frac{-\hat{u}}{\hat{s}-m_{h}^2}, \nb \\
A^S_0&=& \frac{34}{3} \ln^2 \delta_s+ 6 \ln \delta_s \ln
\frac{-\hat{t}}{\hat{s}-m_{h}^2}+ \frac{3}{2} \ln^2
\frac{-\hat{t}}{\hat{s}-m_{h}^2}  \nb \\
&-& \frac{2}{3} \ln \delta_s \ln \frac{-\hat{u}}{\hat{s}-m_{h}^2}-
\frac{1}{6} \ln^2 \frac{-\hat{u}}{\hat{s}-m_{h}^2}-\frac{1}{3}
Li_2[\frac{-\hat{t}}{\hat{s}-m_{h}^2}]+3
Li_2[\frac{-\hat{u}}{\hat{s}-m_{h}^2}].
\end{eqnarray}

\begin{flushleft} {\bf 2.2  Hard collinear gluon emission } \end{flushleft}

In the limit where two of the partons are collinear, the three
body phase space is greatly simplified. And in the same limit, the
leading pole approximation of the matrix element is valid.
According to whether the collinear singularities are initial or
final state in origin, we separate $\hat{\sigma}^{\rm HC}$ into
two pieces
\begin{eqnarray}
\hat{\sigma}^{\rm HC}=\hat{\sigma}_i^{\rm HC}+\hat{\sigma}_f^{\rm
HC}.
\end{eqnarray}
$\hat{\sigma}_i^{\rm HC}$ is the cross section arising from the
case that the emitted gluon is collinear to the initial partons,
$0 \leq t_{15},t_{25} \leq \delta_c \hat{s}$. And
$\hat{\sigma}_f^{\rm HC}$ arises from the case that the emitted
gluon is collinear to the final parton, $0 \leq t_{45} \leq
\delta_c \hat{s}$. We will treat $\hat{\sigma}_f^{\rm HC}$ and
$\hat{\sigma}_i^{\rm HC}$ in Sec. 2.2.1 and Sec. 2.2.2
respectively.

\begin{flushleft} {\bf 2.2.1  Collinear to the final parton} \end{flushleft}

Let $p_4$ and $p_5$ be collinear to each other, $0 \leq t_{45}
\leq \delta_c \hat{s}$. In the collinear limit, the three body
phase space in $d=4-2\epsilon$ time-space dimensions may be
written as\cite{Harris}
\begin{eqnarray}
d\Gamma_3|_{coll}=d\Gamma_2 \frac{(4 \pi)^{\epsilon}}{16 \pi^2
\Gamma(1-\epsilon)}dzd\hat{s}_{45}[\hat{s}_{45} z
(1-z)]^{-\epsilon}.
\end{eqnarray}
The squared matrix element can be factorized as
\begin{eqnarray}
\overline{\sum}|M_f^{\rm HC}(bg \to h^0bg)|^2
\simeq\overline{\sum}|M^0(bg \to h^0b)|^2 ~ P_{bb}(z,\epsilon)
g^2_s\mu_r^{2 \epsilon} \frac{2}{\hat{s}_{45}},
\end{eqnarray}
where $P_{bb}(z,\epsilon)$ is the $d$-dimensional unregulated
($z<1$) splitting function related to the usual Altarelli-Parisi
splitting kernels\cite{Altarelli}. $P_{bb}(z,\epsilon)$ can be
written explicitly as
\begin{eqnarray}
P_{bb}(z,\epsilon)&=&P_{bb}(z)+ \epsilon P'_{bb}(z), \nb \\
P_{bb}(z)&=&C_F \frac{1+z^2}{1-z},~~~~~~~~ P'_{bb}(z)=-C_F (1-z),
\end{eqnarray}
with $C_F=4/3$. After integration over the collinear gluon degrees
of freedom, the cross section $\hat{\sigma}_f^{\rm HC}$ can be
written as\cite{Harris}
\begin{eqnarray}
\label{final collinear cross section} d\hat{\sigma}_f^{\rm
HC}=d\hat{\sigma}^0 \left[\frac{\alpha_s}{2 \pi}
\frac{\Gamma(1-\epsilon)}{\Gamma(1-2 \epsilon)}\left(\frac{4 \pi
\mu_r^2}{\hat{s}}\right)^{\epsilon}\right] \left(\frac{A^{b \to bg
}_1}{\epsilon}+A^{b \to bg}_0 \right),
\end{eqnarray}
where
\begin{eqnarray}
A^{b \to bg }_1&=& C_F(3/2+2\ln\delta_s), \nb \\
A^{b \to bg }_0&=& C_F[7/2-\pi^2/3-\ln^2\delta_s-\ln\delta_c(3/2+2
\ln\delta_s )].
\end{eqnarray}

\begin{flushleft} {\bf 2.2.2  Collinear to the initial parton} \end{flushleft}

Let the hard gluon be emitted collinear to one of the incoming
partons, $0 \leq t_{15} \leq \delta_c \hat{s}_{12}$ or $0 \leq
t_{25} \leq \delta_c \hat{s}_{12}$. In this region, the initial
state partons $i(i=b,g)$ is considered to split into a hard parton
$i'$ and a collinear gluon, $i \to i'g$, with $p_{i'}=zp_i$ and
$k_5=(1-z)p_i$. The matrix element squared for $bg \to h^0bg$
factorizes into the Born matrix element squared and the
Altarelli-Parisi splitting function
\begin{eqnarray}
\overline{\sum}|M_i^{\rm HC}(bg \to h^0bg)|^2 \simeq (4 \pi
\alpha_s \mu_r^{2 \epsilon})\overline{\sum}|M^0(bg \to h^0b)|^2
\left(\frac{-2P_{bb}(z,\epsilon)}{z\hat{t}_{15}}+\frac{-2P_{gg}(z,\epsilon)}{z\hat{t}_{25}}\right),
\end{eqnarray}
where
\begin{eqnarray}
P_{gg}(z,\epsilon)&=&P_{gg}(z)+ \epsilon P'_{gg}(z), \nb \\
P_{gg}(z)&=&2 N[\frac{z}{1-z}+\frac{1-z}{z}+z(1-z)],~~~~~~~~
P'_{gg}(z)=0.
\end{eqnarray}
Using the approximation $p_i-k_5 \simeq zp_i(i=1,2)$, the three
body phase space may be written as
\begin{eqnarray}
d\Gamma_3|_{coll}=d\Gamma_2 \frac{(4 \pi)^{\epsilon}}{16 \pi^2
\Gamma(1-\epsilon)}dzd\hat{t}_{i5}[-(1-z)\hat{t}_{i5}]^{-\epsilon},~~~~(i=1,2).
\end{eqnarray}
Note that the two body phase space should be evaluated at a
squared parton-parton energy of $z\hat{s}$. Therefore, after
integration over the collinear gluon degrees of freedom, we
obtain\cite{Harris}
\begin{eqnarray}
\label{initial collinear}
d\sigma_i^{\rm HC}&=&d\hat{\sigma}^0
\left[\frac{\alpha_s}{2 \pi} \frac{\Gamma(1-\epsilon)}{\Gamma(1-2
\epsilon)}\left(\frac{4 \pi
\mu_r^2}{\hat{s}}\right)^{\epsilon}\right]
(-\frac{1}{\epsilon})\delta_c^{-\epsilon}
[P_{bb}(z,\epsilon)G_{b/A}(x_1/z)G_{g/B}(x_2) \nb \\
&+&P_{gg}(z,\epsilon)G_{g/A}(x_1/z)G_{b/B}(x_2)
+(x_1\leftrightarrow
x_2)]\frac{dz}{z}(\frac{1-z}{z})^{-\epsilon}dx_1dx_2.
\end{eqnarray}
In order to factorize the collinear singularity into the parton
distribution function, we introduce a scale dependent parton
distribution function using the $\overline{\rm MS}$ convention:
\begin{eqnarray}
G_{i/A}(x,\mu_f)=G_{i/A}(x)+(-\frac{1}{\epsilon})\left[\frac{\alpha_s}{2
\pi} \frac{\Gamma(1-\epsilon)}{\Gamma(1-2 \epsilon)}\left(\frac{4
\pi
\mu_r^2}{\mu_f^2}\right)^{\epsilon}\right]\int^1_z\frac{dz}{z}P_{ii}(z)G_{i/A}(x/z),~~~(i=b,g).
\end{eqnarray}
By using above definition, we replace $G_{g,b/A,B}$ in
Eq.(\ref{initial collinear}) and the expression for the initial
state collinear contribution at $O(\alpha_s)$ order is
\begin{eqnarray}
\label{initial collinear cross section} d\sigma_i^{\rm
HC}&=&d\hat{\sigma}^0 \left[\frac{\alpha_s}{2 \pi}
\frac{\Gamma(1-\epsilon)}{\Gamma(1-2 \epsilon)}\left(\frac{4 \pi
\mu_r^2}{\hat{s}}\right)^{\epsilon}\right]\{
\tilde{G}_{g/A}(x_1,\mu_f)G_{b/B}(x_2,\mu_f)+G_{g/A}(x_1,\mu_f)\tilde{G}_{b/B}(x_2,\mu_f)
\nb \\
&+& \sum_{\alpha=g,b}[\frac{A_1^{sc}(\alpha \to \alpha
g)}{\epsilon}+A_0^{sc}(\alpha \to \alpha
g)]G_{g/A}(x_1,\mu_f)G_{b/B}(x_2,\mu_f)+(x_1 \leftrightarrow
x_2)\}dx_1dx_2,
\end{eqnarray}
where
\begin{eqnarray}
A_1^{sc}(b \to bg)&=&C_F(2 \ln \delta_s+3/2), \nb \\
A_1^{sc}(g \to gg)&=&2 N \ln \delta_s + (11 N -2 n_f)/6 , \nb \\
A_0^{sc}&=&A_1^{sc} \ln(\frac{\hat{s}}{\mu_f^2}).
\end{eqnarray}
And
\begin{eqnarray}
\tilde{G}_{\alpha/A,B}(x,\mu_f)=\int^{1-\delta_s}_x
\frac{dy}{y}G_{\alpha/A,B}(x/y,\mu_f)\tilde{P}_{\alpha \alpha}(y),
~~~~(\alpha=g,b),
\end{eqnarray}
with
\begin{eqnarray}
\tilde{P}_{\alpha \alpha}(y)=P_{\alpha \alpha}
\ln(\delta_c\frac{1-y}{y}\frac{\hat{s}}{\mu_f^2})-P'_{\alpha
\alpha}(y,),~~~~(\alpha=g,b).
\end{eqnarray}
We can observe that the sum of the soft (Eq.(\ref{soft cross
section})), collinear(Eq.(\ref{final collinear cross
section}),(\ref{initial collinear cross section})), and
ultraviolet renormalized virtual correction (Eq.(\ref{virtual
cross section})) terms is finite, i.e.,
\begin{eqnarray}
A^S_2&+&A^V_2=0, \nb \\
A^S_1&+&A^V_1+A_1^{b \to bg}+A_1^{sc}(b\to bg)+A_1^{sc}(g\to
gg)=0.
\end{eqnarray}
The final result for the $O(\alpha_s)$ correction consists of two
contributions to the cross section: a two-body term $\sigma^{(2)}$
and a three-body term $\sigma^{(3)}$.
\begin{eqnarray}
\sigma^{(2)}&=&\frac{\alpha_s}{2 \pi} \int dx_1dx_2d\hat{\sigma}^0
\{ G_{g/A}(x_1,\mu_f)G_{b/B}(x_2,\mu_f)[A^S_0+A^V_0+A_0^{b \to
bg}+A_0^{sc}(b\to bg)+A_0^{sc}(g\to gg)] \nb \\
&+&\tilde{G}_{g/A}(x_1,\mu_f)G_{b/B}(x_2,\mu_f)+G_{g/A}(x_1,\mu_f)\tilde{G}_{b/B}(x_2,\mu_f)+(x_1
\leftrightarrow x_2 ) \}.
\end{eqnarray}
And
\begin{eqnarray}
\sigma^{(3)}=\int dx_1dx_2
[G_{g/A}(x_1,\mu_f)G_{b/B}(x_2,\mu_f)+(x_1 \leftrightarrow x_2
)]d\hat{\sigma}^{(3)},
\end{eqnarray}
with the hard-non-collinear partonic cross section given by
\begin{eqnarray}
d\hat{\sigma}^{(3)}=\frac{1}{2\hat{s}_{12}} \int_{\overline{\rm
HC}}\overline{\sum}|M_3(bg \to h^0bg)|^2 d \Gamma_3.
\end{eqnarray}
Finally, the NLO total cross section for $pp$(or $p\bar{p}) \to
bh^0+X$ is
\begin{eqnarray}
\sigma^{NLO}=\sigma^{0}+\sigma^{(2)}+\sigma^{(3)}.
\end{eqnarray}

\section{ NUMERICAL RESULTS AND DISCUSSION}
\par
In the following numerical evaluation, we present the results of
the cross section for the Higgs boson production associated with a
single high-$p_T$ bottom quark via subprocess $bg(\bar{b}g) \to
h^0b(h^0\bar{b})$ at the Fermilab Tevatron and the CERN LHC. At
the LHC, the $b$-jet is required to have a transverse momentum cut
$p_T(b)>30$GeV and a rapidity cut $|\eta(b)|<2.5$. At the
Tevatron, the $b$ tagging regions are taken to be $|\eta(b)|<2$
and $p_T(b)>15$GeV. The SM parameters are taken as: $ m_t=174.3$
GeV, $m_Z = 91.188$ GeV, $m_{W}=80.419~GeV$ and $\alpha_{EW} =
1/128$ \cite{pdg}. For simplicity, the renormalization and
factorization scales are taken as $\mu_r=\mu_f=m_h$. We use the
one-loop formula for the running strong coupling constant
$\alpha_s$ with $\alpha_s(m_Z)=0.117$.
\par
The relevant MSSM parameters in our calculation are: the
parameters $M_{\tilde{Q},\tilde{U},\tilde{D}}$ and $A_{t,b}$ in
squark mass matrices, the higgsino mass parameter $\mu$, the
masses of the gluino $m_{\tilde{g}}$ and the $A^0$ Higgs boson
$m_{A}$, the ratio of the vacuum expectation values of the two
Higgs doublets $\tan\beta$. The squark mass matrix is defined as
\begin{eqnarray}
{\cal M}_{\tilde{q}}^2 &=& \left( \begin{array}{cc} m^2_{\tilde{q}_L} & a_q m_q \\
a_q m_q & m^2_{\tilde{q}_R}
\end{array}\right)
\end{eqnarray}
with
\begin{eqnarray}
m^2_{\tilde{q}_L}&=&M_{\tilde{Q}}^2 + m_q^2 + m_Z^2 \cos 2\beta (I^q_3 - e_q \sin^2 \theta_W), \nb \\
m^2_{\tilde{q}_R}&=&M_{\{\tilde{U},\tilde{D}\}}^2 + m_q^2 + m_Z^2
\cos 2\beta e_q \sin^2 \theta_W   \nb \\
a_q&=&A_q-\mu \{\cot\beta,\tan\beta\},
\end{eqnarray}
for \{up, down\} type squarks. $I_3^q$ and $e_q$ are the third
component of the weak isospin and the electric charge of the quark
$q$. The chiral states $\tilde{q}_L$ and $\tilde{q}_R$ are
transformed into the mass eigenstates $\tilde{q}_{1}$ and
$\tilde{q}_{2}$:
\begin{equation}
\left( \begin{array}{cc} \tilde{q}_{1} \\ \tilde{q}_{2}
\end{array}
\right) = R^{\tilde{q}}\left( \begin{array}{cc} \tilde{q}_L \\
\tilde{q}_R \end{array} \right),~~R^{\tilde{q}} = \left(
\begin{array}{cc}\cos\theta_{\tilde{q}} &
 \sin\theta_{\tilde{q}} \\-\sin\theta_{\tilde{q}} & \cos\theta_{\tilde{q}} \end{array}
 \right).
\end{equation}
Then the mass eigenenvalues $m_{\tilde{q}_{1}}$ and
$m_{\tilde{q}_{2}}$ are given by
\begin{eqnarray}
\left( \begin{array}{cc} m^2_{\tilde{q}_1} & 0 \\
0 & m^2_{\tilde{q}_2}\end{array}\right)=R^{\tilde{q}}{\cal
M}_{\tilde{q}}^2(R^{\tilde{q}})^{\dagger}
\end{eqnarray}
For simplicity, we assume $M_{\tilde{Q}}= M_{\tilde{U}}=
M_{\tilde{D}}=A_t=A_b=600~GeV$, $\mu=200~GeV$, and
$m_A=m_{\tilde{g}}=200~GeV$ by default unless otherwise stated.
\par
In our calculation, we use the program FeynHiggsFast
\cite{Heinemeyer} to generate the mass of Higgs boson($h^0$) and
the mixing angle $\alpha$, and use the CTEQ5L parton distribution
functions\cite{lai}. The $\overline{\rm MS}$ bottom quark mass
$\overline{m}_b$ can be evaluated by using the next-leading order
formula \cite{msbar}. In the following equations, we use
$\overline{m}_b(Q)$ to denote the $\overline{\rm MS}$ bottom quark
mass.
\begin{eqnarray}
\overline{m}_b(Q) &=& U_5(Q,\overline{m}_b)\overline{m}_b(\overline{m}_b),~~~~~~{\rm for}~Q<m_t,  \nb \\
\overline{m}_b(Q) &=&
U_6(Q,m_t)U_5(m_t,\overline{m}_b)\overline{m}_b(\overline{m}_b),~~~~~~{\rm
for}~Q>m_t,
\end{eqnarray}
where $\overline{m}_b=\overline{m}_b(\overline{m}_b)=4.3~GeV$. The
evolution factor $U_f(f=5,6)$ is
\begin{eqnarray}
U_f(Q_2,Q_1)=\left( \frac{\alpha_s(Q_2)}{\alpha_s(Q_1)}\right)^{d^{(f)}}
             [1+\frac{ \alpha_s(Q_1)-\alpha_s(Q_2)}{4 \pi} J^{(f)}], \nb \\
d^{(f)}=\frac{12}{33-2f},~~~~~~~~~~~~~~~J^{(f)}=-\frac{8982-504f+40f^2}{3(33-2f)^2}
\end{eqnarray}
In addition, we also improve the perturbative calculations through
the following replacement\cite{msbar}
\begin{eqnarray}
\overline{m}_b(Q) \to \frac{\overline{m}_b(Q)}{1+\Delta m_b},
\end{eqnarray}
where
\begin{eqnarray}
\Delta m_b = \frac{2 \alpha_s}{3 \pi} m_{\tilde{g}}\tan \beta
I(m_{\tilde{b}_1},m_{\tilde{b}_2},m_{\tilde{g}}),
\end{eqnarray}
with
\begin{eqnarray}
I(a,b,c)=\frac{1}{(a^2-b^2)(b^2-c^2)(a^2-c^2)}(a^2b^2\log
\frac{a^2}{b^2} + b^2c^2\log \frac{b^2}{c^2} + c^2a^2\log
\frac{c^2}{a^2}).
\end{eqnarray}
\par
Fig.6 shows that our NLO-QCD result does not depend on the
arbitrary cutoffs $\delta_s$ and $\delta_c$ by using the two
cutoff phase space slicing method. The two-body($\sigma^{(2)}$)
and three-body($\sigma^{(3)}$) contributions and the NLO cross
section ($\sigma^{NLO}$) are shown as a function of the soft
cutoff $\delta_s$ with the collinear cutoff
$\delta_c=\delta_s/50$, and $\tan\beta=4$ . We can see the NLO
cross section $\sigma^{NLO}$ is independent of the cutoffs. In the
following numerical calculations, we take $\delta_s=10^{-5}$ and
$\delta_c=\delta_s/50$.
\par
Fig.7 shows the dependence of the relative NLO-QCD corrections of
the process $pp$(or $p\bar{p}) \to bg(\bar{b}g) \to
h^0b(h^0\bar{b})+X$ on the mass of $A^0$ Higgs boson $m_A$ at the
LHC. There we take $\tan\beta=10,~40$ and define the relative
correction as
\begin{eqnarray}
\Delta=\frac{\sigma^{NLO}-\sigma^0}{\sigma^0}.
\end{eqnarray}
For $\tan \beta =10$, the relative NLO-QCD correction is about
$58\%$ to $60\%$. And for $\tan \beta =40$, the relative NLO-QCD
correction increases from $66.5\%$ to $73.5\%$ with the increment
of $m_A$ from $200~GeV$ to $800~GeV$.
\par
To compare the production cross section of the processes $pp$(or
$p\bar{p}) \to bg(\bar{b}g) \to h^0b(h^0\bar{b})+X$ in the MSSM
with the corresponding cross section in the SM, we introduce the
ratio of $\sigma_{MSSM}$ and $\sigma_{SM}$. The cross section
$\sigma_{MSSM}$ in the MSSM and the corresponding $\sigma_{SM}$ in
the SM including one-loop order QCD corrections, can be expressed
as
\begin{eqnarray}
\sigma_{MSSM}&=& \sigma^0_{MSSM}(1+\Delta_{MSSM}) \nb \\
\sigma_{SM}&=& \sigma^0_{SM}(1+\Delta_{SM}).
\end{eqnarray}
\par
In Fig.8, we depict the dependence of the ratio $\sigma_{MSSM}
/\sigma_{SM}$ on $A^0$ Higgs boson mass $m_A$ at the Tevatron. We
see that the ratio $\sigma_{MSSM} /\sigma_{SM}$ decreases with the
increment of $m_A$. When $m_A$ is in the range between
$200-300~GeV$, the cross section of the process $p\bar{p} \to
bg(\bar{b}g) \to h^0b(h^0\bar{b})+X$ in the MSSM at the Tevatron
is greatly enhanced over that in the SM, but when $m_A$ increases
from $300~GeV$ to $800~GeV$, the differences of the SM and the
MSSM cross section is becoming smaller. That is because in the
MSSM parameter space for $m_A=200-300~GeV$, the coupling factor
between Higgs boson $h^0$ and bottom quarks
$\sin\alpha/\cos\beta>1$, while in the region of $m_A>300~GeV$,
the value of $\sin\alpha/\cos\beta$ is getting close to 1.
\par
Fig.9 shows the cross sections of the process $pp \to bg(\bar{b}g)
\to h^0b(h^0\bar{b})+X$ at the LHC and $p\bar{p} \to bg(\bar{b}g)
\to h^0b(h^0\bar{b})+X$ at the Tevatron as the functions of
$\tan{\beta}$. We scale the mass of the Higgs boson $m_h$ under
the X-axis to show the changes of the phase space with
$\tan{\beta}$. The figure shows that when $\tan{\beta}$ varies
from $5$ to $10$, the cross sections drop quickly and $m_{h^0}$
increases faster than in other $\tan{\beta}$ regions. But when
$\tan{\beta}$ is larger than $10$, the variation of mass of Higgs
boson($h^0$) is not sensitive to $\tan{\beta}$, and the cross
sections drop smoothly with the increasing of $\tan{\beta}$. When
$\tan{\beta}$ approaches to the value of 50, the relative NLO-QCD
corrections take the value about $70\%$ at the LHC and $85\%$ at
the Tevatron. The curve behavior shown in Fig.9, comes from the
feature of the phase space variation versus $\tan{\beta}$. In
Fig.10 we present the dependence of $\sigma_{MSSM} /\sigma_{SM}$
on $\tan{\beta}$ at the LHC and Tevatron. We find that the cross
section of the processes $pp$(or $ p\bar{p}) \to bg(\bar{b}g) \to
h^0b(h^0\bar{b})+X$ in the MSSM is greatly enhanced compared with
that in the SM in all the plotted range of $\tan{\beta} = 4 \sim
50$ with the parameter space we take. That is due to the stronger
coupling strength between Higgs boson $h^0$ and bottom quarks with
large $\tan{\beta}$ in the MSSM than in the SM. Then the
production of the processes $pp$(or $p\bar{p}) \to
h^0b(h^0\bar{b})+X$ at hadron colliders could be enhanced
significantly in the MSSM.
\par
To study the decoupling behavior of SUSY QCD correction, we assume
all SUSY mass related parameters have the same quantity and are
pushed to a large value except $\mu$ is taken to be $200GeV$. We
denote $M_{\tilde{Q}}= M_{\tilde{U}}=
M_{\tilde{D}}=A_t=A_b=m_A=m_{\tilde{g}}$ collectively by $M_s$ and
define the relative SUSY QCD correction as
\begin{eqnarray}
\Delta_{SQCD}=\frac{\delta \sigma^{SQCD}}{\sigma^0},
\end{eqnarray}
where $\delta \sigma^{SQCD}$ is the cross section correction
contributed by the SUSY QCD diagrams shown in Fig.3. In the mass
region of $m_A=M_s>500~GeV$, $\sin\alpha/\cos\beta$ is close to 1.
Therefore, with this parameter choice, the dependence of $\Delta
\sigma_{SQCD}$ on $M_s$ can demonstrate the decoupling behavior of
SUSY QCD correction in the processes $pp$(or $p\bar{p}) \to
h^0b(h^0\bar{b})+X$ at hadron colliders. In Fig.11, we depict the
relative SUSY QCD correction $\Delta_{SQCD}$ as the functions of
$M_s$ at the LHC. From the figure, we can see that for
$\tan\beta=10$ the relative SUSY QCD correction is approaching to
zero with $M_s$ increasing to $2.2~TeV$, but for $\tan\beta=40$
the relative correction decreases to about $4\%$ with $M_s$
increasing to $3~TeV$. That shows obviously that the large
$\tan\beta$ enhances the SUSY QCD corrections and delays the
decoupling feature.
\par
To analyze the scale dependence of the cross sections, we
introduce the ratio of the cross section at scale $Q$ to the cross
section at scale $Q=m_{h^0}$ and depict the $\sigma(Q) /
\sigma(Q=m_{h^0})$ as a function of $Q/m_{h^0}$ at the Tevatron in
Fig.12. The scale variation of the NLO-QCD cross section may be
serves as an estimate of the remaining theoretical uncertainty of
the high order corrections. Fig.12 shows that it is evident that
the one-loop NLO-QCD corrections reduce the LO scale dependence.
\par
In summary, we have computed the production of Higgs boson $h^0$
associated with a single high-$p_T$ bottom quark via subprocess
$bg(\bar{b}g) \to h^0b (h^0\bar{b})$ including the NLO-QCD
corrections in the MSSM at the CERN LHC and the Fermilab Tevatron.
We find that due to the enhancement of the Yukawa couplings
strength of the down-type quarks with Higgs bosons at large
$\tan{\beta}$, the cross section of the $pp \to h^0b(h^0\bar{b})$
can reach hundreds of fermi barn at the LHC and the cross section
of the $p\bar{p} \to h^0b(h^0\bar{b})$ can reach dozens of fermi
barn at the Tevatron. The NLO-QCD correction reaches $50\% \sim
70\%$ at the LHC and $60\% \sim 85\%$ at the Tevatron in the
parameter space we have chosen.
\par
\noindent{\large\bf Acknowledgments:} The authors are very
grateful to Prof. T. Han and Dr. Wan Lang-Hui for their
encouragement and the useful discussions. This work was supported
in part by the National Natural Science Foundation of China and a
grant from the University of Science and Technology of China.
\par
\begin{center} {\bf Appendix A} \end{center}
\par
\setcounter{equation}{0}
\renewcommand{\theequation}{A.\arabic{equation}}
\par
In Appendix A, we list the explicit expressions of the
renormalization constants. By using the $\overline{\rm MS}$ scheme
the renormalization constants defined in Eq.(\ref{defination of
renormalization constants}) are expressed as:
\begin{eqnarray}
\delta Z_{b}&=& -\frac{\alpha_s}{4 \pi} C_F \Delta, \nb \\
\frac{\delta m_b}{m_b} &=& -\frac{\alpha_s}{2 \pi}  C_F \Delta,  \nb  \\
\delta Z_g &=& -\frac{\alpha_s}{4 \pi} (\frac{1}{2} N- \frac{1}{2}
\beta_0) \Delta,
\end{eqnarray}
where $\Delta=\frac{1}{\epsilon}-\gamma_E+ \log(4 \pi)$,
$\beta_0=(11 N - 2 n_f)/3$, $N=3$ and $C_F=4/3$.
\par
We define the following symbols
\begin{eqnarray}
C_{gb \bar{b}}
&=& \frac{1}{2} \delta Z_g +\delta Z_b + \delta Z_{g_s},  \nb \\
C_{hb\bar{b}}&=&\delta Z_b+ \delta Y_b,~~~~~~~ C_{b\bar{b}}=\delta
Z_b.
\end{eqnarray}
Then the amplitude from the diagrams which contain counter terms
$M^{CT}$ can be written as
\begin{eqnarray}
M^{CT}&=&
\frac{g_sY_b}{\hat{s}^2}\bar{u}_i(k_4)[\hat{s}(C_{gb\bar{b}}+C_{hb\bar{b}})(\rlap/p_1
+ \rlap/p_2)- C_{b\bar{b}} (\rlap/p_1 + \rlap/p_2)^2]
\gamma_{\mu} u_j(p_1)\epsilon_{\mu}(p_2)T^a_{ij}  \nb \\
&+&
\frac{g_sY_b}{\hat{t}^2}\bar{u}_i(k_4)[\hat{t}(C_{gb\bar{b}}+C_{hb\bar{b}})(\rlap/p_1
- \rlap/k_3)- C_{b\bar{b}} (\rlap/p_1 - \rlap/k_3)^2] \gamma_{\mu}
u_j(p_1)\epsilon_{\mu}(p_2)T^a_{ij}.
\end{eqnarray}

\begin{center} {\bf Appendix B} \end{center}
\par
\setcounter{equation}{0}
\renewcommand{\theequation}{B.\arabic{equation}}
In Appendix B, we present expressions of the non-vanishing form
factors in Eq.(\ref{virtual diagrams amplitude}). We denote the
couplings between gluino , quark and squark as
\begin{eqnarray}
\tilde{g}-\tilde{t}_i-\bar{t}:~(V^{(1)}_{\tilde{g}\tilde{t}_it}P_L
+V^{(2)}_{\tilde{g}\tilde{t}_it}P_R)T^a,~~~~~~\tilde{g}-\tilde{b}_i-\bar{b}:~(V^{(1)}_{\tilde{g}\tilde{b}_ib}P_L
+V^{(2)}_{\tilde{g}\tilde{b}_ib}P_R)T^a,
\end{eqnarray}
where
\begin{eqnarray}
V^{(1)}_{\tilde{g}\tilde{t}_{1}t,\tilde{g}\tilde{b}_{1}b}=-i\sqrt{2}g_s
\sin{\theta}_{\tilde{t},\tilde{b}},~~~~~
V^{(1)}_{\tilde{g}\tilde{t}_{2}t,\tilde{g}\tilde{b}_{2}b}=i\sqrt{2}g_s\cos{\theta}_{\tilde{t},\tilde{b}}, \nb \\
V^{(2)}_{\tilde{g}\tilde{t}_{1}t,\tilde{g}\tilde{b}_{1}b}=-i\sqrt{2}g_s\cos{\theta}_{\tilde{t},\tilde{b}},~~~~~
V^{(2)}_{\tilde{g}\tilde{t}_{2}t,\tilde{g}\tilde{b}_{2}b}=-i\sqrt{2}g_s\sin{\theta}_{\tilde{t},\tilde{b}}.
\end{eqnarray}
And the couplings between the Higgs boson and two quarks(squarks)
are denoted as
\begin{eqnarray}
h-b-\bar{b}:~~V_{hbb},~~~~~~h-\tilde{b}_i-\tilde{b}_j^{\dagger}:~V_{h
\tilde{b}_i \tilde{b}_j},
~~~~~~h-\tilde{t}_i-\tilde{t}_j^{\dagger}:~V_{h \tilde{t}_i
\tilde{t}_j},
\end{eqnarray}
where the explicit expressions of the $V_{hbb}$,
$V_{h\tilde{b}_i\tilde{b}_j}$ and $V_{h\tilde{t}_i\tilde{t}_j}$
can be found in Ref.\cite{vertex}. For simplicity, we introduce
the following abbreviations,
\begin{eqnarray}
V^{(1)*}_{\tilde{g}\tilde{b}_ib} \cdot
V^{(1)}_{\tilde{g}\tilde{b}_jb}&=&F^{1}_{ij},~~~~~
V^{(1)*}_{\tilde{g}\tilde{b}_ib} \cdot V^{(2)}_{\tilde{g}\tilde{b}_jb}=F^{2}_{ij},  \nb \\
V^{(2)*}_{\tilde{g}\tilde{b}_ib} \cdot
V^{(1)}_{\tilde{g}\tilde{b}_jb}&=&F^{3}_{ij},~~~~~
V^{(2)*}_{\tilde{g}\tilde{b}_ib} \cdot
V^{(2)}_{\tilde{g}\tilde{b}_jb}=F^{4}_{ij},
\end{eqnarray}
and
\begin{eqnarray}
B_0^{a},B_1^{a} &=& B_0,B_1[p_1, 0, 0], ~~~~
B_0^{b},B_1^{b} = B_0,B_1[k_3-p_1, 0, 0], \nb \\
B_0^{c},B_1^{c} &=& B_0,B_1[-p_1-p_2, 0, 0],~~~~
B_0^{d},B_1^{d} = B_0,B_1[k_3 - p_1, m_{\tilde{g}}, m_{\tilde{b}_i}], \nb \\
B_0^{e},B_1^{e} &=& B_0,B_1[-p_1 - p_2, m_{\tilde{b}_i},
m_{\tilde{g}}],~~~~
C_0^{a},C_{kl}^{a} = C_0,C_{kl}[-k_3, p_1, 0, 0, 0], \nb \\
C_0^{b},C_{kl}^{b} &=& C_0,C_{kl}[-p_2, -p_1, 0, 0, 0],~~~~
C_0^{c},C_{kl}^{c} = C_0,C_{kl}[-p_2, k_3 - p_1, 0, 0, 0], \nb \\
C_0^{d},C_{kl}^{d} &=& C_0,C_{kl}[k_3, -p_1 - p_2, 0, 0, 0],~~~~
C_0^{e},C_{kl}^{e} = C_0,C_{kl}[-k_3, p_1, m_{\tilde{b}_j}, m_{\tilde{b}_i}, m_{\tilde{g}}], \nb \\
C_0^{f},C_{kl}^{f} &=& C_0,C_{kl}[-p_2, -p_1,  m_{\tilde{g}},
m_{\tilde{g}}, m_{\tilde{b}_i}],~~~~
C_0^{g},C_{kl}^{g} = C_0,C_{kl}[-p_2, -p_1, m_{\tilde{b}_i}, m_{\tilde{b}_i}, m_{\tilde{g}}], \nb \\
C_0^{h},C_{kl}^{h} &=& C_0,C_{kl}[-p_2, k_3 - p_1, m_{\tilde{g}},
m_{\tilde{g}}, m_{\tilde{b}_i}],~~~~
C_0^{m},C_{kl}^{m} = C_0,C_{kl}[-p_2, k_3 - p_1, m_{\tilde{b}_i}, m_{\tilde{b}_i}, m_{\tilde{g}}], \nb \\
C_0^{n},C_{kl}^{n} &=& C_0,C_{kl}[k_3, -p_1 - p_2,
m_{\tilde{b}_j}, m_{\tilde{b}_i}, m_{\tilde{g}}],~~~~
D_0^{a},D_{kl}^{a} = D_0,D_{kl}[-p_2, k_3, -p_1, 0, 0, 0, 0], \nb \\
D_0^{b},D_{kl}^{b} &=& D_0,D_{kl}[-p_2, k_3, -p_1,
m_{\tilde{b}_j},
m_{\tilde{b}_j},m_{\tilde{b}_i}, m_{\tilde{g}}], \nb \\
D_0^{c},D_{kl}^{c} &=& D_0,D_{kl}[k_3, -p_1, -p_2,
m_{\tilde{b}_j},m_{\tilde{b}_i}, m_{\tilde{g}},  m_{\tilde{g}}]
\end{eqnarray}
Since we neglect the bottom quark mass throughout the calculation
except in the Yukawa couplings, the form factors
$f_{1,2,7,8,9,10}$ are irrelevant to our results and we do not
present them here.
\par
For diagrams shown in Fig.2(SM-like QCD corrections), we can write
the form factors as
\begin{eqnarray}
f_3&=&f_4=\frac{3 g_s V_{hbb}}{16 \pi^2}[C^c_{22} - C^c_{23} +(
D^a_{12}- 2 D^a_{11}  - 2 D^a_{21}   + D^a_{24})
m_h^2 + (D^a_{13} + D^a_{25}) \hat{s}\nb \\
&-& (D^a_{12}  - D^a_{13} + D^a_{24}  - D^b_{25}) \hat{t}]
\end{eqnarray}
\begin{eqnarray}
f_5&=&f_6=\frac{-g_s V_{hbb}}{96 \pi^2 \hat{t}}[26 + 4 B^b_{0} -
32 B^a_0 - 16 B^b_{1} - 120 C^c_{24}+ 16(2 C^a_{11}
 -  C^a_{12}) m_h^2 \nb \\
  &+& (2 C^a_{11} - 9 C^c_{0}  - 4 C^a_{0}  -
18 C^c_{11} + 9 C^c_{12} + 16 C^a_{12}  + 40 C^c_{22}
   - 40 C^c_{23}
+ 74 D^a_{27} ) \hat{t}\nb \\
&+& (18 D^a_{12} - 29 D^a_{11}+ 2 D^a_{13} - 29 D^a_{21} - 6
D^a_{22}  + 18 D^a_{24} + 2 D^a_{25} + 4 D^a_{26}) m_h^2 \hat{t}
\nb \\
&+& ( 16 D^a_{13}- 16 D^a_{12}   + 2 D^a_{22} + 6 D^b_{24} + 16
D^a_{25}  - 13 D^a_{26}
) \hat{s} \hat{t}\nb \\
&+& ( 16 D^a_{13}- 16 D^a_{12} + 2 D^a_{22}+ 4 D^a_{24} - 16
D^a_{24} + 16 D^a_{25}- 4 D^a_{26}) \hat{t}^2]
\end{eqnarray}
\begin{eqnarray}
f_{11}&=&f_{12}=\frac{g_s V_{hbb}}{192 \pi^2 \hat{s}} [(24 - 32
B^c_{0} - 16 B^c_{1} - 104 C^b_{24} + 16(2 C^d_{11}-
C^d_{12})m_h^2)(1+\hat{s}/\hat{t}) \nb \\
&+& (16 D^a_{13}-18 D^a_{12} - 2 D^a_{13} + 9
D^a_{23} + 9 D^a_{25} - 18 D^a_{26}) \hat{s} \nb \\
&+& (18 C^b_{0} -18 C^c_{0}  - 9 C^c_{11} + 9 C^b_{11} - 7
C^c_{12}  - 16 C^d_{12} + 16 C^a_{12}  + 16 C^c_{22} - 16 C^c_{23}
+ 16 C^b_{23} + 54 D^a_{27})  \nb \\
&+&( 36 D^a_{12} - 27 D^a_{11} - 5 D^a_{13}+ 9 D^a_{24}  - 18 D^a_{25}  + 9 D^a_{26}) m_h^2 \nb \\
&+&(16 D^a_{13}- 36 D^a_{12} - 2 D^a_{13}
 + 9 D^a_{23} - 9 D^a_{24}  + 9
D^a_{25} - 9 D^a_{26})\hat{t}]
\end{eqnarray}
\par
For diagrams shown in Fig.3(SUSY QCD corrections), we can write
the form factors as, (The summation over the particle indices of
squark $(i,j=1,2)$ has been omitted.)
\begin{eqnarray}
f_3&=&\frac{g_s}{96 \pi^2}[(-C^m_{22} + C^m_{23}) F^4_{ii} V_{hbb}
+ (D^b_{0} + D^b_{11} + D^b_{12})
m_{\tilde{g}}F^3_{ji}V_{h\tilde{b}_j\tilde{b}_i}] \nb \\
f_4&=&f_3(F^4_{ii} \to F^1_{ii} ~~F^3_{ji} \to F^2_{ji}) \nb \\
f_5&=&\frac{g_s}{96 \pi^3 \hat{t}}[(B^d_0 - 8 B^d_1 - 18 C^h_{24}
- 2 C^m_{24} + (9 C^h_{22}  + C^m_{22}  - 9 C^h_{23} -
C^m_{23})\hat{t} ) F^4_{ii} V_{hbb}\nb \\
&+& (8 C^e_{0} - (D^b_{0} + 9 D^c_{0} + D^b_{11} + 9 D^c_{11} +
D^c_{12}) \hat{t})
m_{\tilde{g}}F^3_{ji}V_{h\tilde{b}_j\tilde{b}_i}] \nb \\
f_6&=&f_5(F^4_{ii} \to F^1_{ii} ~~F^3_{ji} \to F^2_{ji}) \nb \\
f_{11}&=&\frac{g_s}{384 \pi^2 \hat{s} \hat{t}}[\hat{t} (9 + 16
B^e_0 + 16 B^e_1 - 36 C^f_{24} + 4 C^g_{24} + 18 C^f_{12}\hat{s} +
18 C^f_{23}\hat{s} - 18 C^f_{0} m_{\tilde{g}}^2) F^1_{ii}
V_{hbb}\nb \\
&+& \hat{s} (9 + 16 B^d_0 + 16 B^d_1 - 36 C^h_{24} + 4 C^i_{24} +
18 C^h_{22} \hat{t} - 18 C^h_{23} \hat{t} - 18 C^h_{0}
m_{\tilde{g}}^2)F^4_{ii} V_{hbb} \nb\\
&-& 2 (8 C^e_{0} \hat{s} + 8 C^n_{0} \hat{t} - 9 D^c_{0}
\hat{s}\hat{t}) m_{\tilde{g}}
F^3_{ji}V_{h\tilde{b}_j\tilde{b}_i}]\nb \\
f_{12}&=&f_{11}(F^4_{ii} \leftrightarrow F^1_{ii} ~~F^3_{ji}
\leftrightarrow F^2_{ji})
\end{eqnarray}
\par
In our paper we adopt the definitions of the one-loop integrals in
Ref. \cite{s13}. The numerical calculation of the vector and
tensor one-loop integral functions can be traced back to four
scalar loop integrals $A_{0}$, $B_{0}$, $C_{0}$, $D_{0}$ as shown
in Ref.\cite{passvelt}. Many of the integral functions contain the
soft and collinear IR singularities, the formulas to calculate
these integrals can be found in Ref. \cite{beenakker}.

\vskip 10mm
\begin{flushleft} {\bf Figure Captions} \end{flushleft}

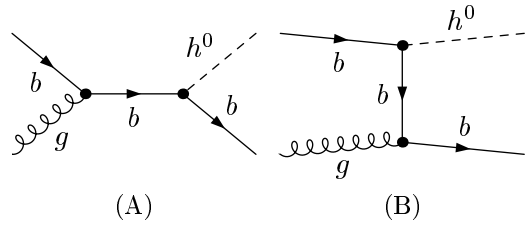
{\bf Fig.1} Leading order Feynman diagrams for the subprocess of
$bg \to h^0b$.
\par
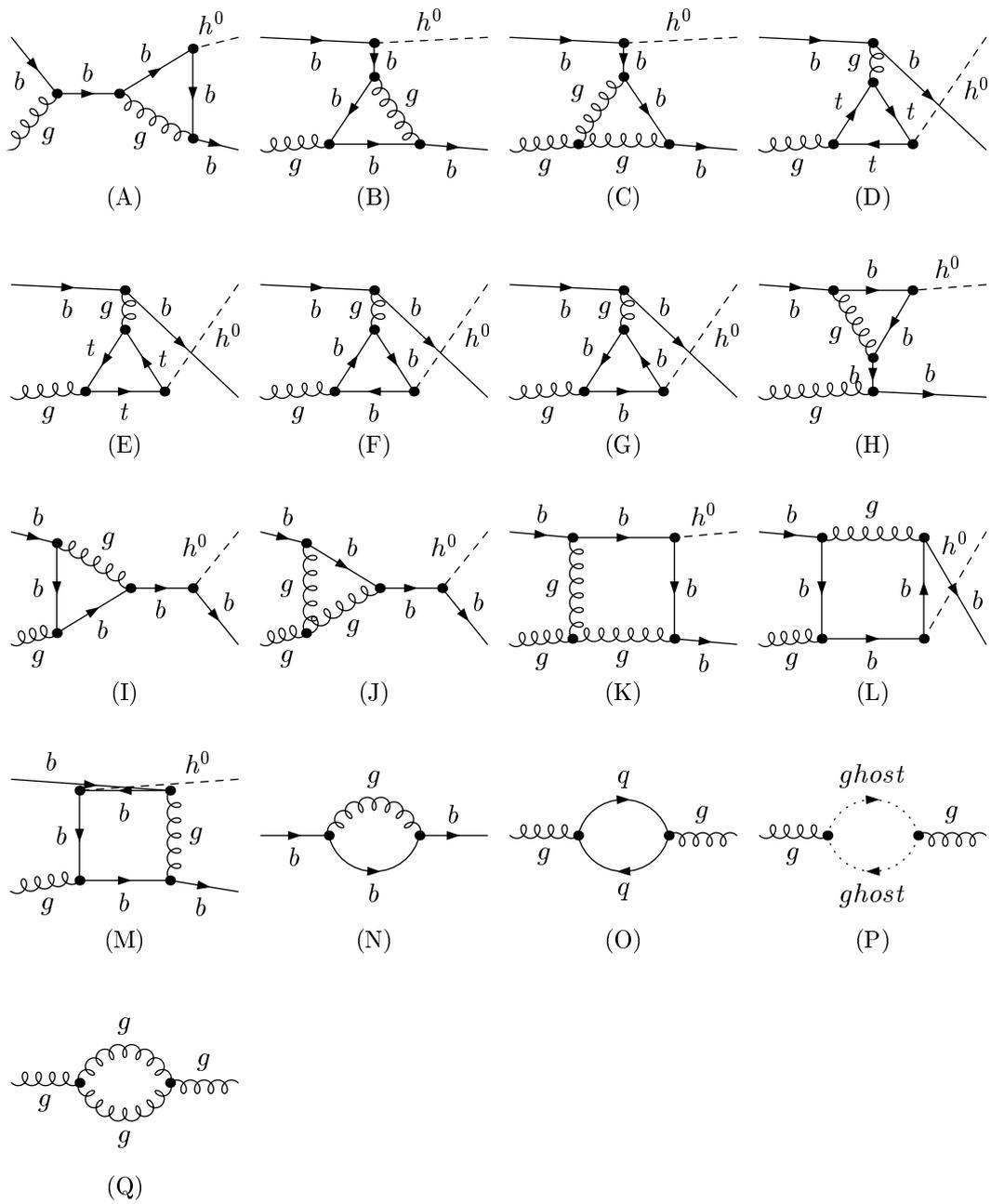
{\bf Fig.2} Virtual one-loop Feynman diagrams of the SM-like QCD
corrections.
\par
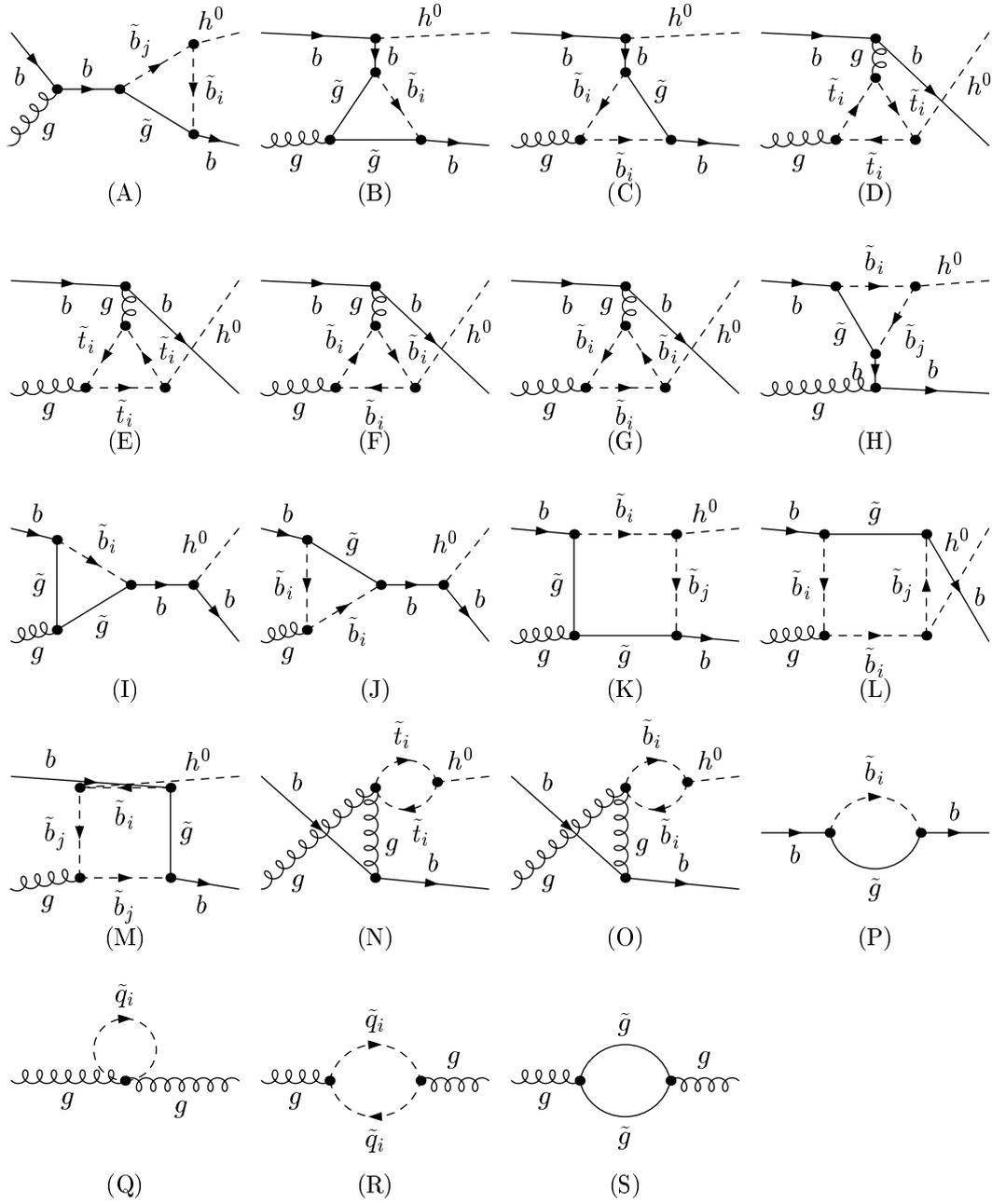
{\bf Fig.3} Virtual one-loop Feynman diagrams of the SUSY QCD
corrections.
\par
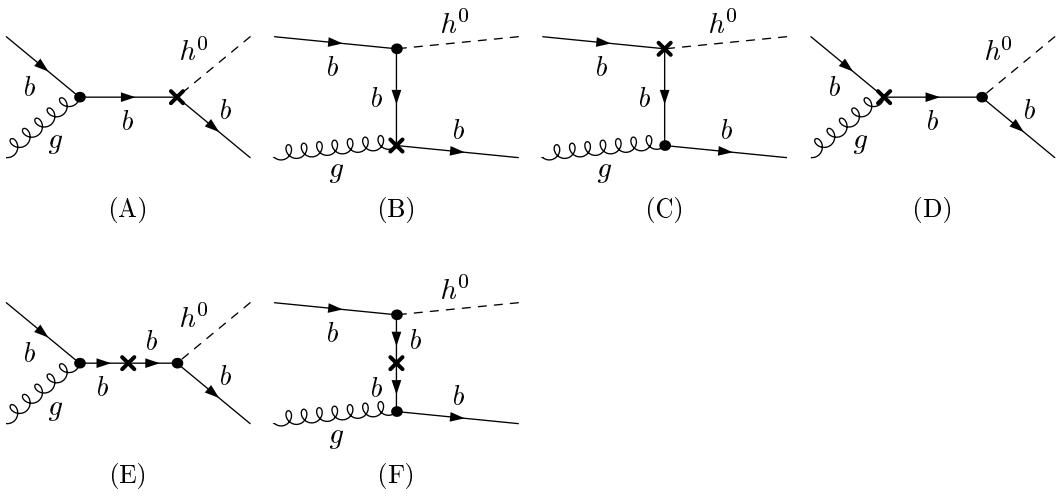
{\bf Fig.4}  Feynman diagrams that contain counter-term.
\par
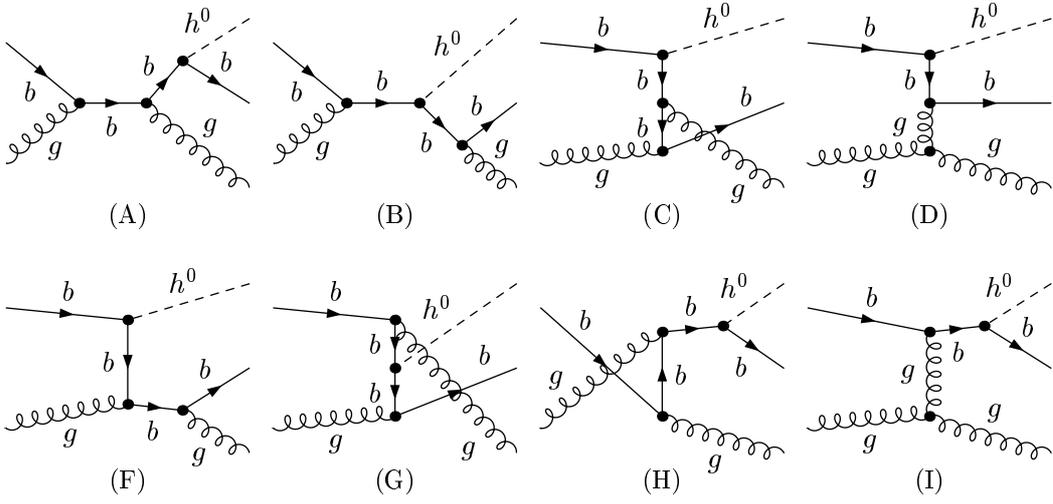
{\bf Fig.5} Feynman diagrams for the real gluon emission.
\par
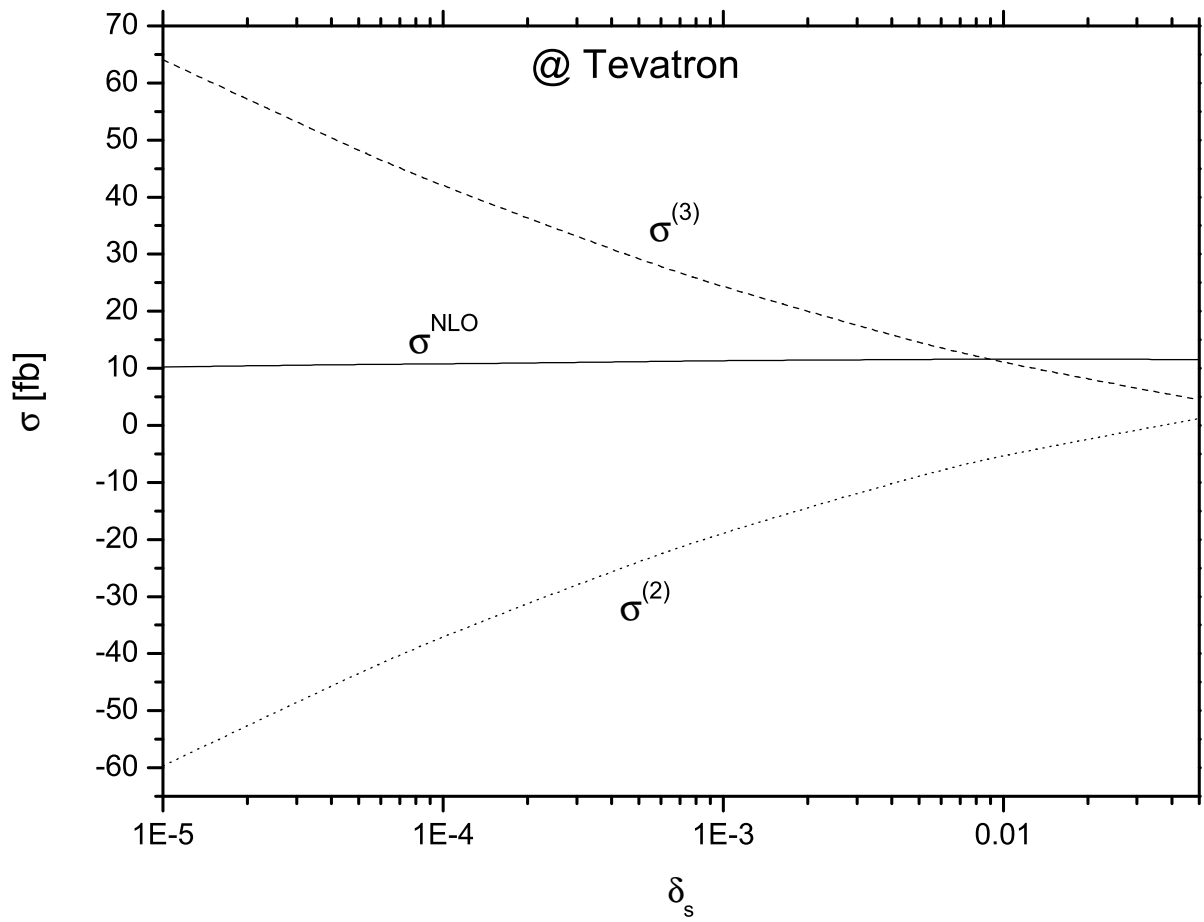
{\bf Fig.6} Dependence of the cross sections for the $h^0b$
production at the Tevatron on the cutoff $\delta_s$ with
$\delta_c=\delta_s/50$.
\par
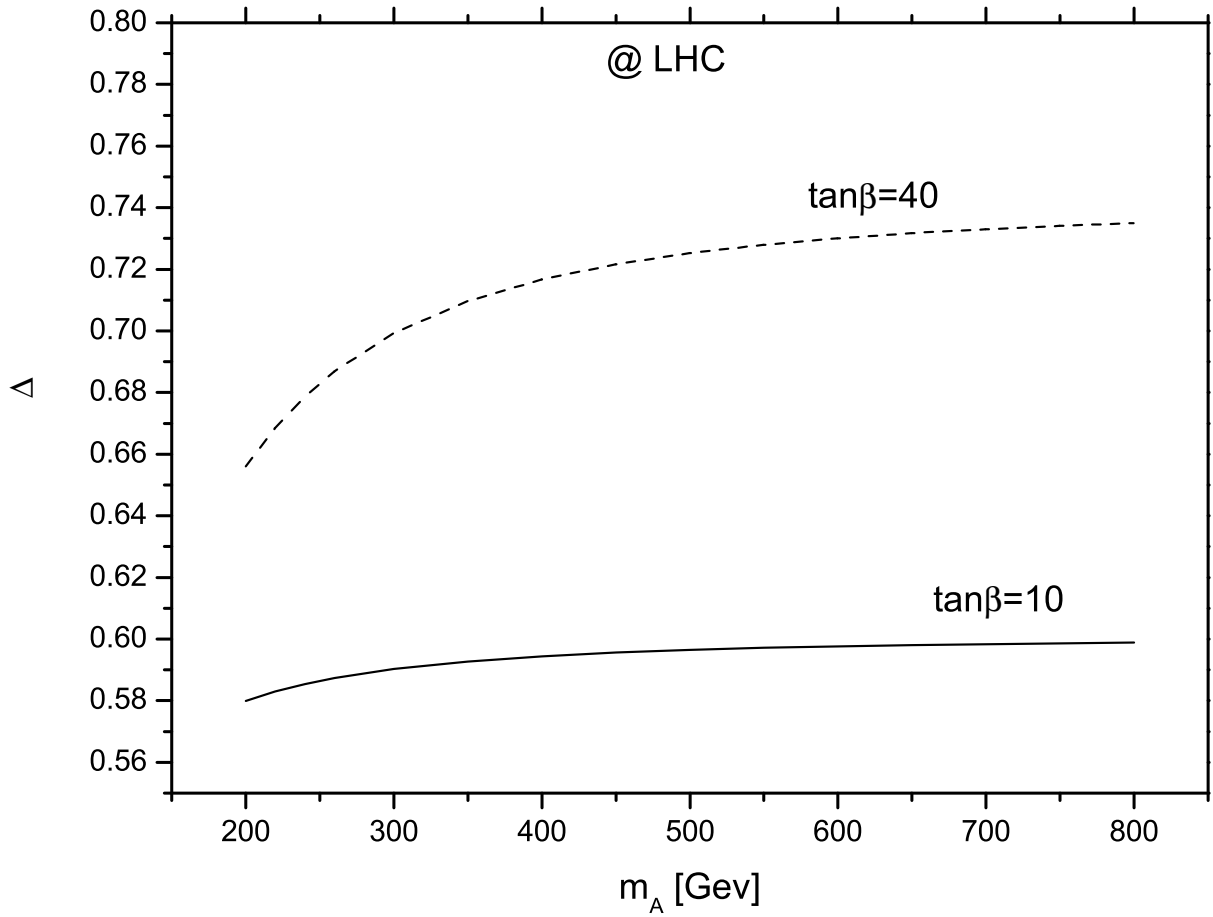
{\bf Fig.7} The dependence of the relative NLO-QCD correction on
the $m_A$ with $\tan\beta=4,10$ at the LHC.
\par
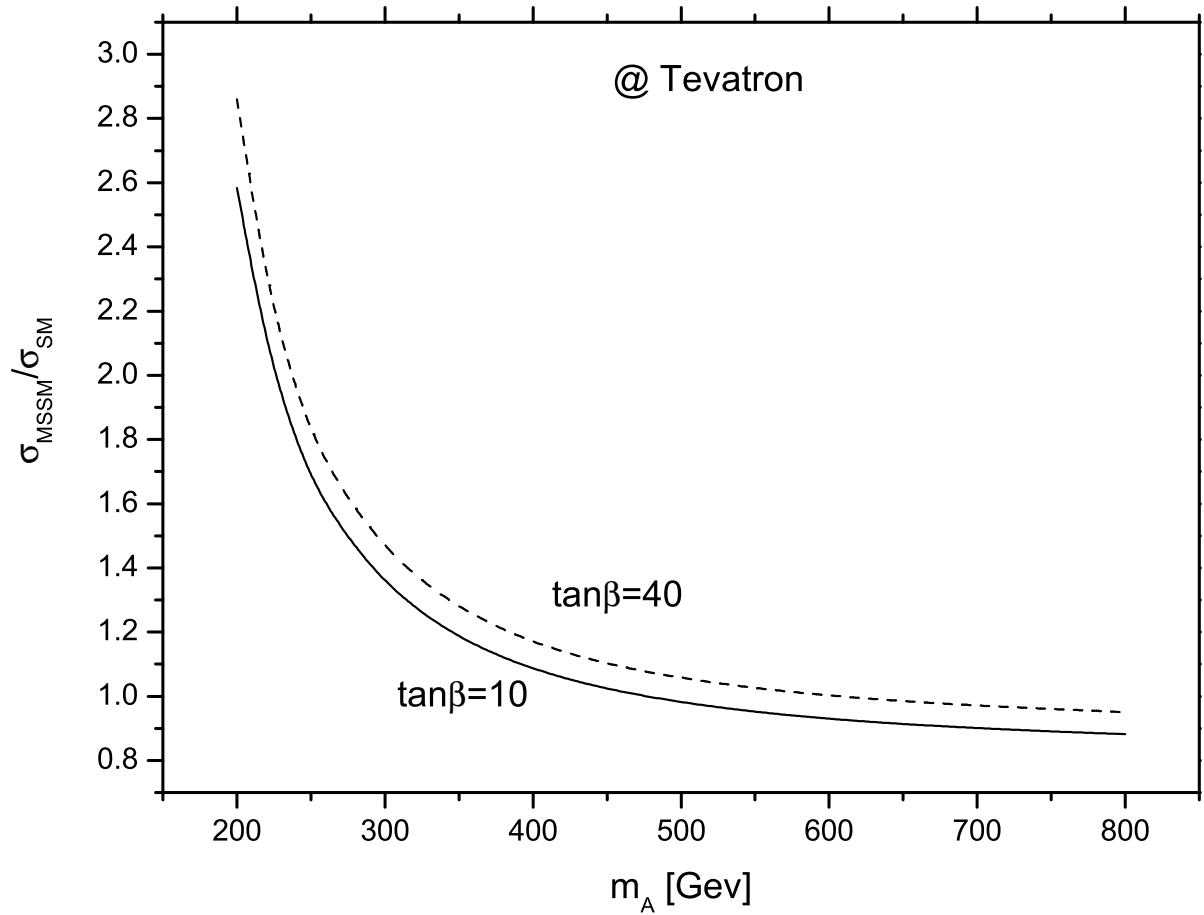
{\bf Fig.8} The dependence of $\sigma_{MSSM} /\sigma_{SM}$ on the
$m_A$ with $\tan\beta=4,10$ at the Tevatron.
\par
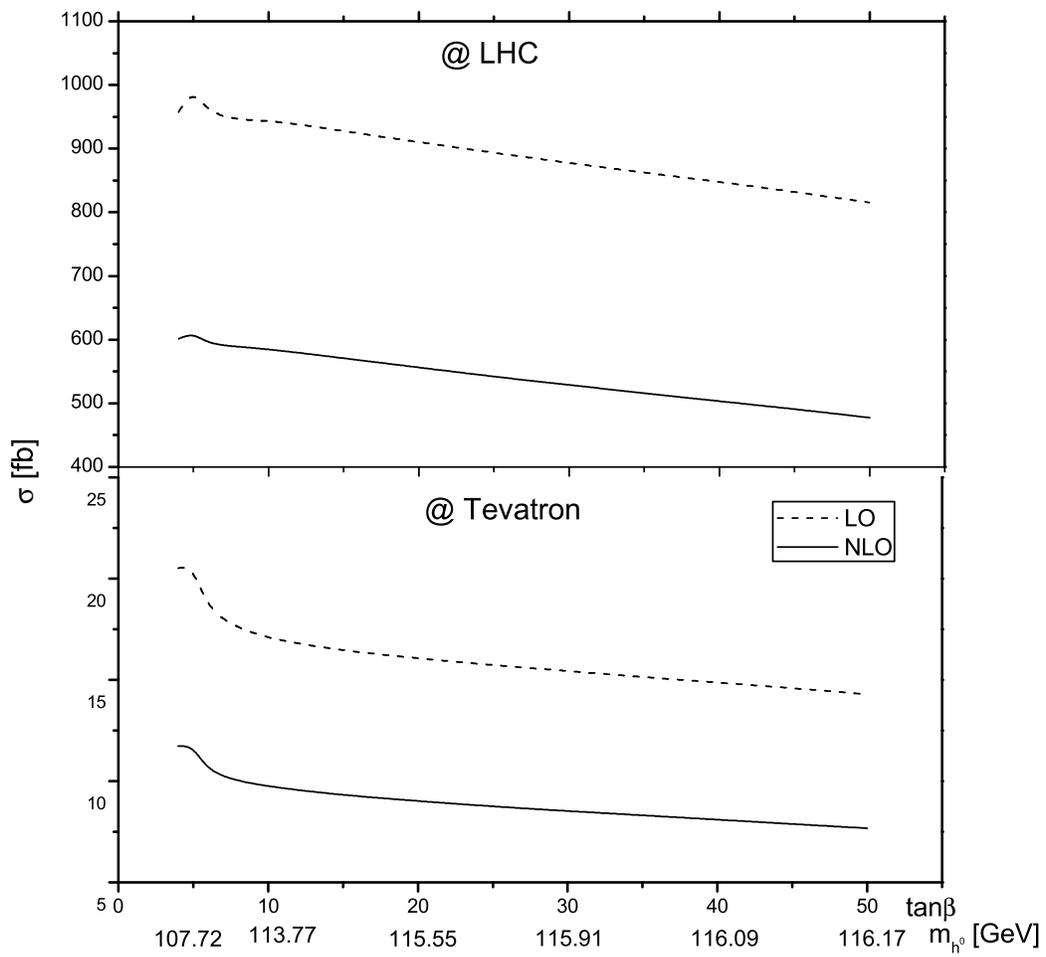
{\bf Fig.9} The dependence of the cross sections
 on $\tan\beta$ at the LHC and the Tevatron.
\par
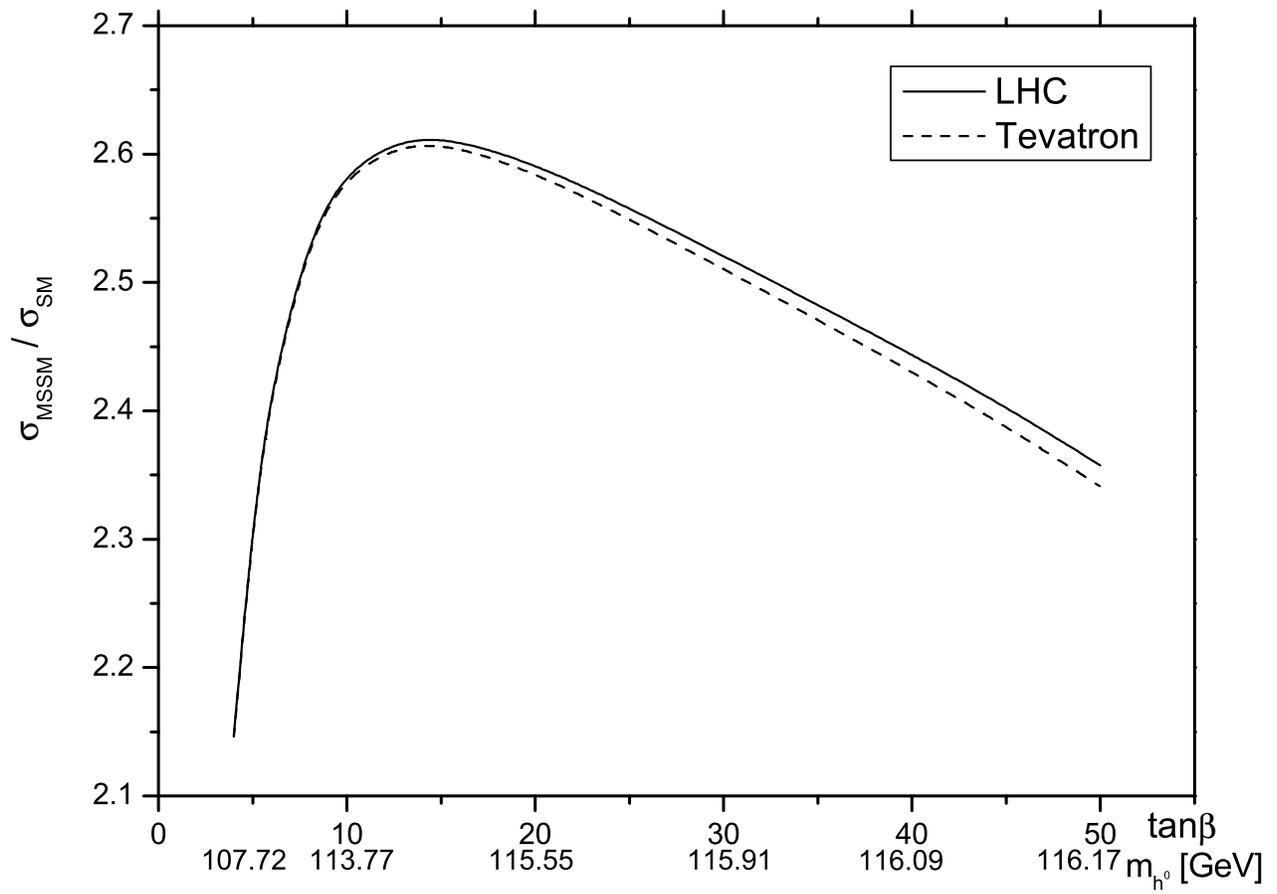
{\bf Fig.10} The dependence of $\sigma_{MSSM} /\sigma_{SM}$
 on $\tan\beta$ at the LHC and the Tevatron.
\par
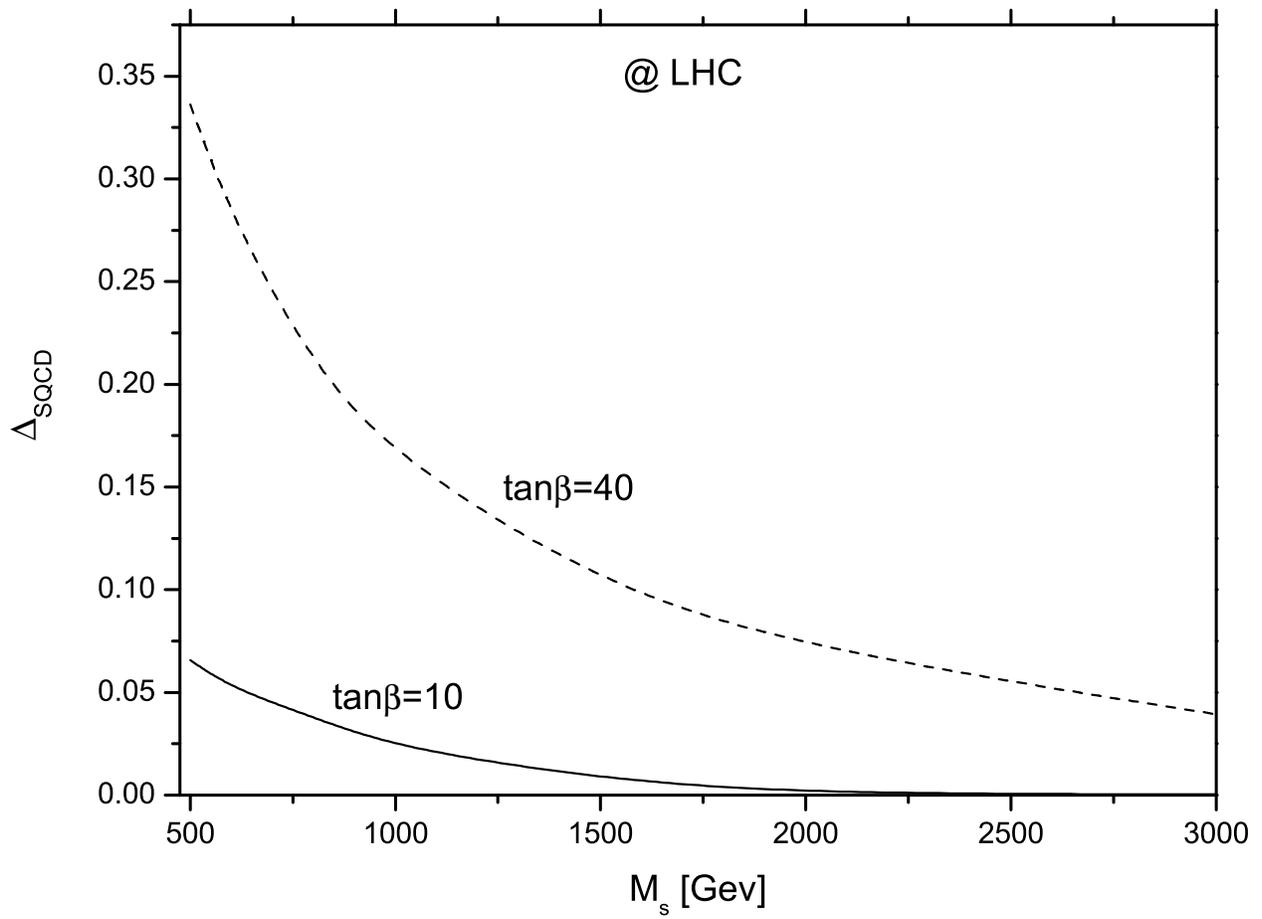
{\bf Fig.11} The dependence of $\Delta_{SQCD}$
 on $M_s$ at the LHC.
\par
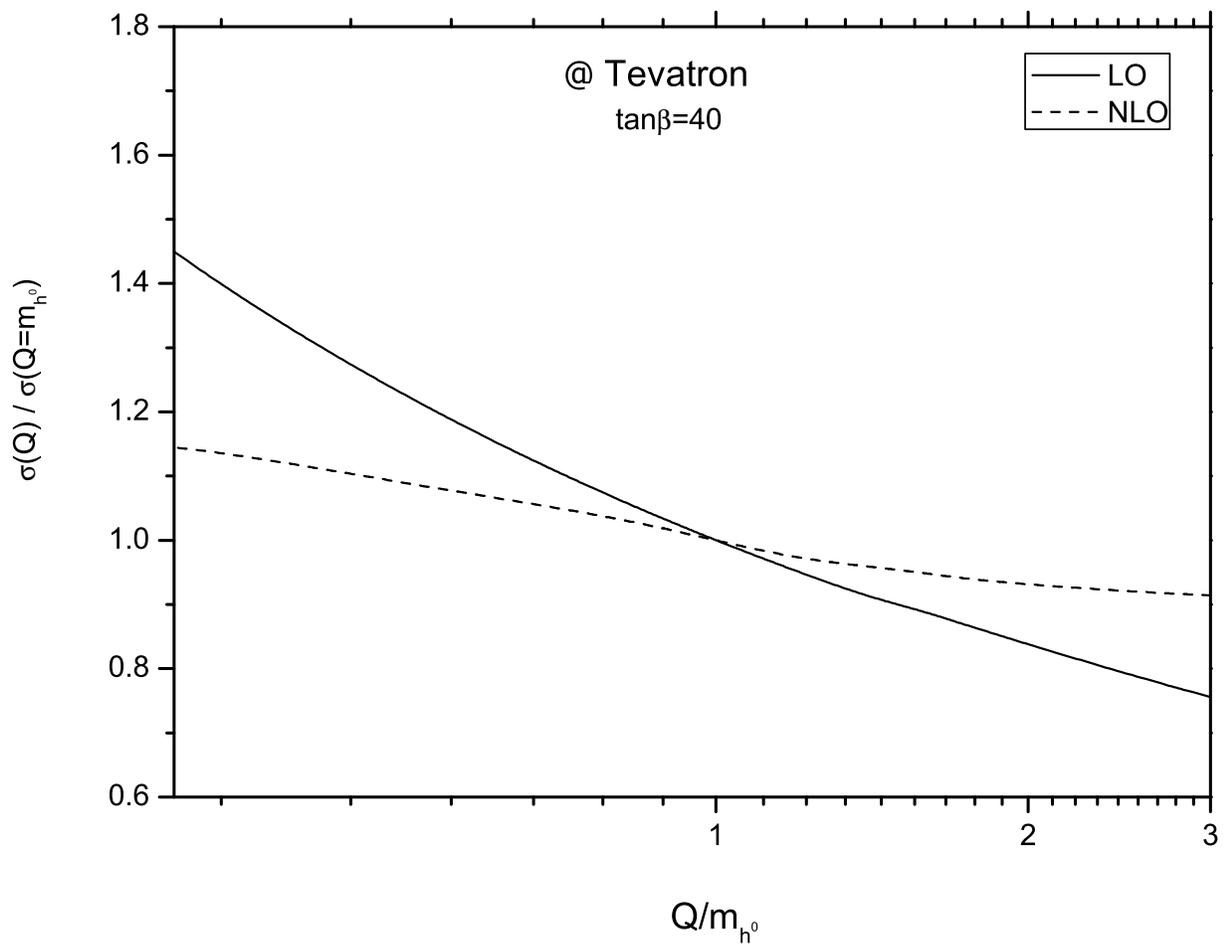
 {\bf Fig.12} The
variation of the $\sigma(Q) / \sigma(Q=m_{h^0})$ with the ratio
$Q/m_{h^0}$ at the Tevatron.
\end{document}